\documentclass[12pt]{iopart}
\usepackage{graphicx}
\usepackage{color}
\newcommand{\red}[1]{#1}
\def\beq{\begin{equation}}
\def\eeq{\end{equation}}
\def\bea{\begin{eqnarray}}
\def\eea{\end{eqnarray}}
\def\beax{\begin{eqnarray*}}
\def\eeax{\end{eqnarray*}}
\def\half{\frac{1}{2}}
\begin{document}
\title[Quantum and Classical Statistical Mechanics \dots]{Quantum and Classical Statistical Mechanics of a Class of non-Hermitian Hamiltonians}
\author{H F Jones$^1$ and E S Moreira, Jr.$^2$}
\address{$^1$Physics Department, Imperial College, London SW7 2AZ, UK}
\address{$^2$Instituto de Ci\^encias Exatas, Universidade Federal de Itajub\'a,
 37500-903 Itajub\'a, MG, Brazil}
\eads{\mailto{h.f.jones@imperial.ac.uk}, \mailto{moreira@unifei.edu.br}}

\begin{abstract}
This paper investigates the thermodynamics of a large class of
non-Hermitian, $PT$-symmetric oscillators, whose energy
spectrum is entirely real. The spectrum is estimated by
second-order WKB approximation, which turns out to be very
accurate even for small quantum numbers, and used to generate the
quantum partition function. Graphs showing the thermal behavior of
the entropy and the specific heat, at all regimes of temperature,
are given. To obtain the corresponding classical partition
function it turns out to be necessary in general to integrate over a complex ``phase
space". For the wrong-sign quartic, whose equivalent Hermitian
Hamiltonian is known exactly, it is demonstrated explicitly how
this formulation arises, starting from the Hermitian case.
\end{abstract}
\section{Introduction}
\label{introduction}
\red{The recent surge of} interest in Hamiltonians which, although non-Hermitian, nevertheless have a completely real spectrum
began with the pioneering paper of Bender and Boettcher~\cite{BB}, which gave strong numerical
and analytical evidence that the spectrum of the class of Hamiltonians
\bea\label{M=1}
H = p^2 -(ix)^N
\eea
was completely real and positive for $N \ge 2$, and attributed this reality to the (unbroken)
$PT$ symmetry of the Hamiltonian. Since then a large number of $PT$ -symmetric models
have been explored (see, e.g.,~\cite{LZ}), and it was found that the phenomenon is rather general.

The natural metric arising in such theories is not positive definite, which precludes a straightforward physical interpretation in terms of probability amplitudes. However, it turns out to be possible to
construct~\cite{BBJ} a grading operator $C$, which gives a positive-definite metric $\eta_+=PC$. In contrast to standard
quantum mechanics, where the metric is the same for any Hermitian theory, here the metric is determined in each case by the Hamiltonian itself. In the Schr\"odinger wave-function formulation of quantum mechanics, this metric appears explicitly in the calculation of Green functions; however, in the path-integral of functional formulation its role is much more subtle~\cite{JR}.

The class of Hamiltonians (\ref{M=1}) can be thought of as a continuation in the exponent of the potential, starting with the harmonic oscillator, $N=2$. A more general class of Hamiltonians~\cite{BBM,BBJS} is obtained by continuation off the $x^{2M}$ oscillator:
\begin{equation}\label{M<>1}
 H=p^{2}+x^{2M}(ix)^{\epsilon},
\end{equation}
where $M=1,2,3,\ldots$; $\epsilon\geq 0$. They include the
harmonic oscillator $H=p^{2}+x^{2}$ ($M=1$, $\epsilon=0$),
and non-Hermitian Hamiltonians such as
$H=p^2+ix^3$  ($M=1$, $\epsilon=1$),
$H=p^2-x^4$ ($M=1$, $\epsilon=2$) and
$H=p^2-x^6$ ($M=2$, $\epsilon=2$).
The eigenvalue equation corresponding to (\ref{M<>1}) is
\beq -\psi_{n}''(x)+x^{2M}(ix)^{\epsilon}\psi_{n}(x)=E_{n}\psi(x)
\label{sequation}
\eeq (throughout the text, $\hbar=k_{B}=1$), where
$\psi_{n}(x)$ is required to vanish as
$|x|\rightarrow\infty$. As explained in \cite{BBM, BBJS}, when the total exponent
$N\equiv2M+\epsilon$
is greater than 4 the boundary condition can no longer be satisfied with $x$ real, and
one is obliged to analytically continue the eigenvalue equation (\ref{sequation}) into the complex plane.
Specifically it should be continued into the lower half $x$ plane
within a Stokes wedge symmetrically placed with respect to the imaginary axis~\footnote{this is why
the superficially Hermitian Hamiltonians $H=p^2-x^4$ and $H=p^2-x^6$ are in fact non-Hermitian and
$PT$-symmetric.}. When this is done, the energy
spectrum turns out to be real, discrete and positive~\cite{BB,BBM,BBJS}. A rigorous proof of this property
was eventually constructed by Dorey et al.~\cite{DDT}. \red{For reviews of the whole field of non-Hermitian
Hamiltonians see \cite{rev}.}

In principle it is possible~\cite{AM02} to relate
$H$ by a similarity transformation to an equivalent Hermitian Hamiltonian $h$ with the
same energy spectrum. However, such a programme is difficult to implement,
and an exact form for $h$ is available only in very few cases.
It should be stressed that, unlike (\ref{sequation}), the eigenvalue equation
corresponding to $h$ can always be solved on the real axis.

\red{Possible physical applications of these unusual Hamiltonians are beginning to emerge. We mention
here the ``quantum brachistochrone", in which the standard lower bound for the transition time between
two states can be circumvented~\cite{QB1} by a judicious interplay of Hermitian and non-Hermitian systems.
How this could be achieved in practice has been recently discussed in \cite{QB2}. In analogue optical
situations, where the refractive index plays the role of the potential, PT-symmetric systems have been shown
to have unusual and interesting properties~\cite{DC1, DC2}.}

\red{Perhaps surprisingly, little attention has been paid so far to the thermodynamics of such systems. In the present
paper we investigate the thermodynamics of the Hamiltonians of (\ref{M<>1}), which we hope will act as a template
for the investigation of other non-Hermitian systems. As far as the quantum partition thermodynamics is concerned
we need only calculate the energy levels, on which the partition function depends. The high-temperature limit
then probes the classical mechanics of these systems, and we use }\red{this to investigate how the classical partition function
should be defined, which is not obvious {\sl a priori}.}

The paper is organized as follows. In Sec.~\ref{spectrum} we show that the energies $E_{n}$ of
(\ref{sequation}) can be estimated to a high degree of accuracy, even for small values of $n$,
using the WKB approximation including the subleading
contribution.  In Sec.~\ref{qmechanics} we use the energy levels so determined to
evaluate numerically the canonical
partition function, which then yields plots illustrating
the thermal behavior of the entropy and the specific heat,
at low, intermediate and high temperatures. The plots also
show how the parameters $M$ and $\epsilon$ affect these
quantities. In the high-temperature limit, still using the  WKB formula for $E_{n}$,
we obtain a closed formula for the semiclassical partition
function by integration over $n$.

This is then used, in Sec.~\ref{cmechanics}, as a tool for investigating how the semiclassical partition
function $Z_{cl}$ associated with $H$ in (\ref{M<>1}) should be expressed as
a ``phase-space" integral. In general this must differ from the usual real phase-space integral,
which does not converge.
The special case $H=p^{2}-x^4$ (the wrong-sign quartic oscillator)
is ideal for testing these ideas, because in this case one has an explicit form for
the equivalent isospectral Hermitian Hamiltonian $h$. Thus one can start by calculating
the semiclassical partition function using the real phase space associated with $h$, and
then transform the variables to find the correct integral expression for $Z_{cl}$ in terms of the
variables appearing in $H$. Sec.~\ref{disc} includes a summary and further discussion.
\section{The WKB approximation for the spectrum}
\label{spectrum}
An extremely accurate approximation for the spectrum of $H$
in (\ref{M<>1}) can be obtained using the WKB method \red{including the subleading
contribution. In \cite{BB,BBM} the spectra were so calculated for $M=1$.
This section generalizes those results to the case $M > 1$.}

One begins by solving the equation
$x^{2M}(ix)^{\epsilon}=E$ to determine the turning points,
\beq x_{\pm}=E^{1/N}e^{-i\pi(1/2\mp M/N)},
\label{tpoints}
\eeq recalling that $N=2M+\epsilon$.
It should be pointed out that  $x_{\pm}$
are in the wedge mentioned in the previous section.
For a given $M$, when $\epsilon=0$ the turning points are
$x_{\pm}=\pm E^{1/N}$. As $\epsilon$
increases from zero, these points migrate from the real axis towards the
negative imaginary axis of the complex $x$ plane.

The leading WKB contribution is obtained
by imposing the condition
\beq (n+1/2)\pi=\int_{x_{-}}^{x_{+}}dx\sqrt{E-x^{2M}(ix)^{\epsilon}}
\label{wkb1}
\eeq over a path for which the integration is real.
Choosing the path as the ray going from $x_{-}$ to $0$,
followed by that from $0$ to $x_{+}$, (\ref{wkb1})
can be recast as
\beq (n+1/2)\pi=2\sin(M\pi/N)E^{1/2+1/N}\int_{0}^{1}ds\sqrt{1-s^{N}},
\label{wkb2}
\eeq which yields the first factor in (\ref{wkbspectrum}) below.

To calculate the subleading WKB contribution,
instead of (\ref{wkb1}), one uses~\cite{BBM, BO}
\beq\label{wkb3}
(n+1/2)\pi=(1/2i)\oint_{C} dx Q^{1/2}
+(1/2i)\oint_{C} dxQ''/48Q^{3/2},
\eeq
where $Q(x):=x^{2M}(ix)^{\epsilon}-E$ and the
contour $C$ encircles counterclockwise the rays
used to calculate the integration in (\ref{wkb1}).
It follows then that
\beax
\frac{1}{2i}\oint_{C} dx\frac{Q''(x)}{48Q^{3/2}(x)}&=&
\frac{1}{24}N(N-1)\sin(M\pi/N)\\
&& \times E^{-1/2-1/N}\int_{0}^{1}ds\frac{s^{N-2}}{(1-s^{N})^{3/2}},
\eeax
which when added to (\ref{wkb2}) yields (for large $n$)
\bea\label{wkbspectrum}
E_{n}&=&
\left(\frac{\Gamma(3/2+1/N)\sqrt{\pi}(n+1/2)}{\sin(M\pi/N)\Gamma(1+1/N)}\right)^{2N/(N+2)}\nonumber\\
&&\hspace{2cm}\times \left(1+\frac{N(N-1)\sin^2(M\pi/N)\cot(\pi/N)}{3\pi (n+1/2)^2 (N+2)^2}\right),
\eea
after identities involving products of
gamma functions are used.

The energy spectrum corresponding to the Hamiltonian $p^{2}+|x|^{N}$\red{, which
we consider for purposes of comparison, }is obtained~\cite{BBM}
by omitting the factors $\sin(M\pi/N)$ and $\sin^{2}(M\pi/N)$ in (\ref{wkbspectrum})
[corresponding to $M=N/2$],
and the results in \cite{BB,BBM} are reproduced by setting $M=1$ in (\ref{wkbspectrum}).
Tables 1 and 2 show that (\ref{wkbspectrum}) is an excellent approximation for both the Hermitian
and non-Hermitian versions even for small
quantum numbers.
\begin{table}[h]
\caption{
Energy levels $E_n$ of the potential $|x|^4$. The successive columns are
$n$, first- and second-order WKB, the numerical results from \cite{ban78},
and the approximation from \cite{tur79}.}
\begin{indented}
\item[]\begin{tabular}{lllll}
\br
$n$ & \mbox{WKB}1 & \mbox{WKB}2 & \mbox{Exact} & \mbox{Turschner}\\
\hline
0 & 0.86714532 & 0.98982129 & 1.06036209 & 1.032458\\
1 & 3.75191992 & 3.81089637 & 3.79967303 & 3.785676 \\
10 & 50.2401523 & 50.2562691 & 50.2562545 & 50.254484 \\
50 & 407.868707 & 407.874363 & 407.874363 & 407.87365 \\
100 & 1020.986417 & 1020.989992 & 1020.989992 & 1020.989538\\
\br
\end{tabular}
\end{indented}
\end{table}

\begin{table}[h]
\caption{
Energy levels $E_n$ of the potential $-x^4$. The successive columns are
$n$, first- and second-order WKB, and the numerical results from \cite{BB}.}
\begin{indented}
\item[]
$\begin{tabular}{llll}
\br
$n$ & \mbox{WKB}1 & \mbox{WKB}2 & \mbox{Exact} \\
\hline
0 & 1.37651 & 1.47388 & 1.4771\\
1 & 5.9558 & 6.00261 & 6.0033 \\
2 & 11.769 & 11.8023 & 11.8023\\
3 & 18.4321 & 18.4588 & 18.4590 \\
\br
\end{tabular}$
\end{indented}
\end{table}

\section{Quantum Statistical mechanics}
\label{qmechanics}
In quantum statistical mechanics the partition function
is given by
\beq Z(T)=\sum_{n=0}^{\infty}e^{-E_{n}/T}.
\label{pfunction}
\eeq Note that this is exactly the same as it would be for a Hermitian Hamiltonian, with no appearance
of the metric operator. As was first pointed out by Jakubsk\'y~\cite{Jak}, this is due to the cyclic property
of the trace in $Z=\Tr{e^{-H/T}}$.

For the $PT$-symmetric oscillators we are considering, we use the second-order WKB approximation
of (\ref{wkbspectrum}) to evaluate the energies $E_{n}$.  This is guaranteed to be accurate at high temperatures when the populations of states with high energies are appreciable, so that the relevant
values of $n$ are large. However, as we have illustrated above, the energies so obtained are
very accurate, even for small $n$, so the corresponding approximation for $Z$ should also be accurate at
low temperatures.

\begin{figure}[thp]
\begin{center}
\includegraphics[scale=0.8]{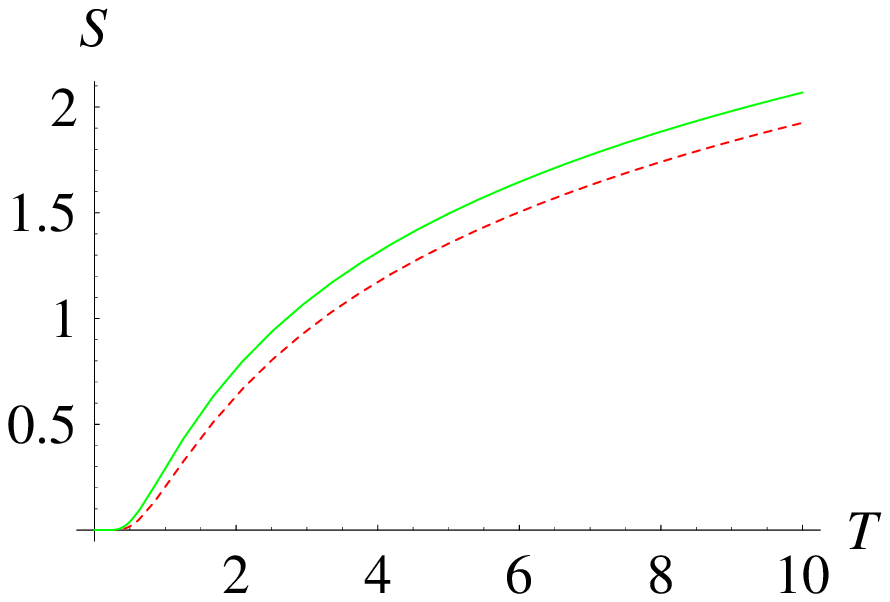}
\includegraphics[scale=0.8]{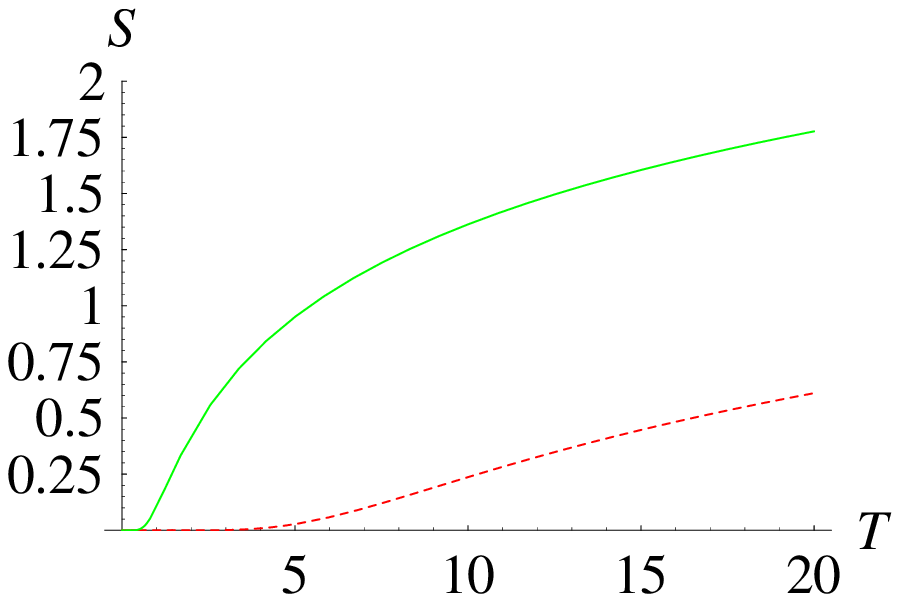}
\caption{$S$ versus $T$ for $N$ =3 (top) and 10 (bottom), using
the second-order WKB approximation (\ref{wkbspectrum}) for the energies.
In each case the solid curve (green) corresponds to
$V=|x|^N$ and the dashed curve (red) to $V=x^2(ix)^{N-2}$ (
$M=1$, $\epsilon=N-2$).
\label{Splot}}
\end{center}
\end{figure}

\begin{figure}[thp]
\begin{center}
\includegraphics[scale=0.8]{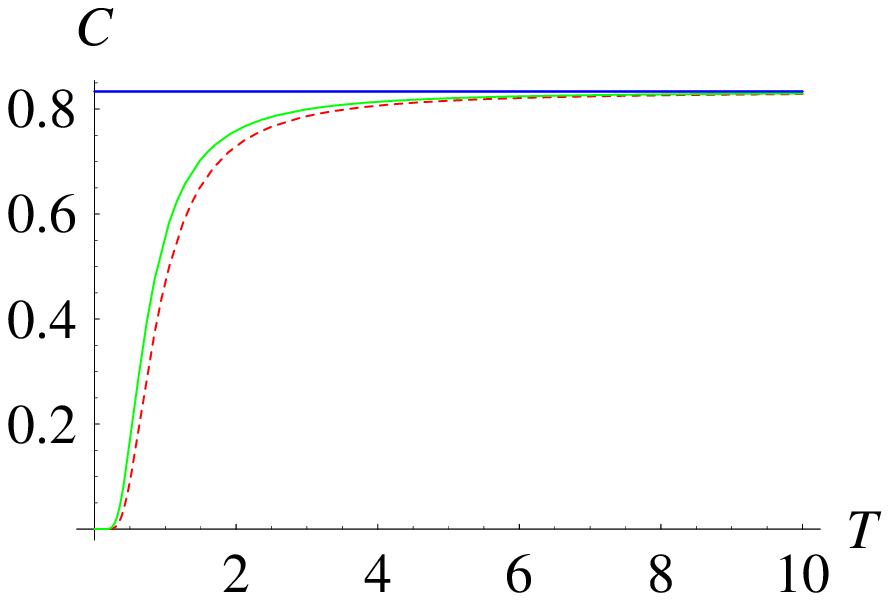}
\includegraphics[scale=0.8]{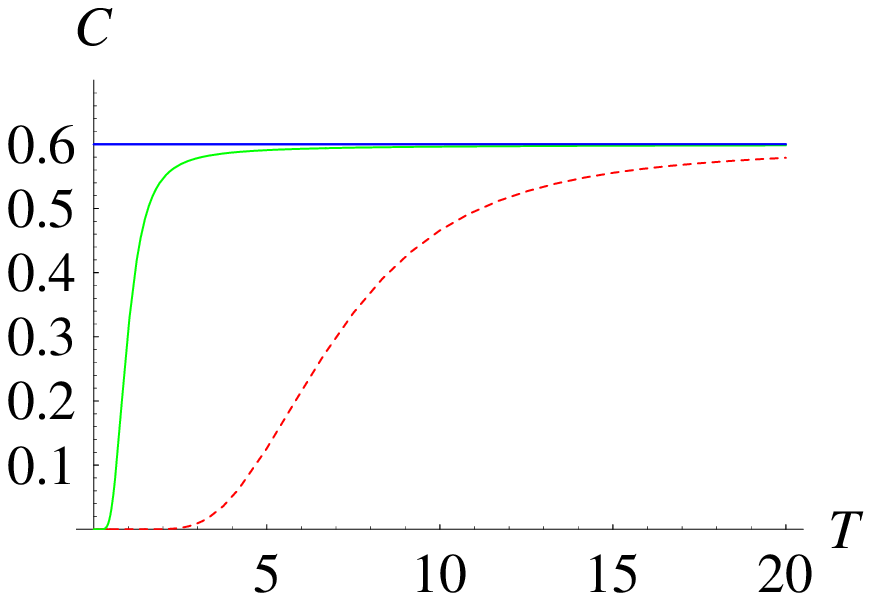}
\caption{$C$ versus $T$ for $N$=3 (top) and 10 (bottom), using
the second-order WKB approximation (\ref{wkbspectrum}) for the energies.
In each case the solid curve (green) corresponds to
$V=|x|^N$ and the dashed curve (red) to $V=x^2(ix)^{N-2}$ (
$M=1$, $\epsilon=N-2$).
\label{Cplot}}
\end{center}
\end{figure}

Once we have calculated $Z(T)$ we may evaluate other thermodynamic quantities in the standard way from the
free energy $F=-T\log Z$, which yields the entropy $S=-dF/dT$, the internal energy $U=F+TS$, and the
specific heat $C=dU/dT$. In Figures~\ref{Splot} and \ref{Cplot} we illustrate
the thermal behavior of $S$ and $C$ for $M=1$ and various values of
$\epsilon$, and compare them with those obtained for the Hermitian potential $|x|^N$.

Apart from the case where $M=1$ and $\epsilon=0$ (the harmonic oscillator), there is no closed form
for the summation in (\ref{pfunction}). However, a closed form can be obtained
at high temperatures, where as  mentioned above, (\ref{wkbspectrum})
is most reliable. Working with the leading contribution in (\ref{wkbspectrum}),
since the subleading contribution is suppressed at large $n$, we obtain
\begin{equation}
Z(T)=\sum_{n=0}^{\infty}e^{-\beta E_{0}(2n+1)^{2N/(N+2)}},
\label{htpfunction}
\end{equation}
where $E_{0}$ denotes the first factor in (\ref{wkbspectrum})
when $n=0$, and $\beta:=1/T$. As $\beta E_{0}\rightarrow 0$, the summation can
be replaced by an integration,
\begin{equation}
Z_{cl}(T)=\int_{0}^{\infty}dn\ e^{-(\Theta/T)(n+1/2)^{2N/(N+2)}},
\label{zfunction}
\end{equation}
where $\Theta$ is the characteristic temperature
of the oscillator,
\begin{equation}
\Theta:=\left[\frac{\Gamma(3/2+1/N)\sqrt{\pi}}{\sin(M\pi/N)\Gamma(1+1/N)}\right]^{2N/(N+2)}.
\label{ctemperature}
\end{equation}
In fact, $\beta E_{0}\rightarrow 0$ means $\Theta/T\ll 1$; i.e., one is dealing
with the classical regime, as indicated by the change in notation from $Z$ to $Z_{cl}$.
A few manipulations in (\ref{zfunction}) lead to
\begin{equation}
Z_{cl}(T)=\Gamma (3/2+1/N)\left(\frac{T}{\Theta}\right)^{1/2+1/N}.
\label{scpf}
\end{equation}
By simple inspection, (\ref{scpf}) yields
the well known expression $T/\Theta$ for the harmonic oscillator $N=2$.

$Z_{cl}(T)$ in (\ref{scpf}) leads to the entropy
\begin{equation}
S=[1/2+1/N]\left[\log (T/\Theta)+1\right]+\log \Gamma(3/2+1/N)
\label{entropy}
\end{equation}
and specific heat
\begin{equation}
 C=1/2+1/N,
\label{sheat}
\end{equation}
corresponding to the (classical) thermal behavior
on the right-hand part of the plots.
The characteristic temperature $\Theta$ in (\ref{ctemperature})
gives the magnitude of the energy gap separating the first excited
state and the ground state. As (\ref{ctemperature}) clearly shows,
for a given $M$, $\Theta$ is an unbounded increasing function of $N$.
Since $\Theta$ separates the classical thermal behavior
[(\ref{entropy}) and (\ref{sheat})]
from the quantum thermal behavior (corresponding in the plots
to the  drop toward zero as $T\rightarrow 0$),
one sees that the larger $N$ the higher the temperature
up to which the quantum behavior still prevails.
These features are clear in the graphics.

It should be noted that
\beq Z_{cl}(T)=Q_{cl}(N,T) \sin(M\pi/N),
\label{zq}
\eeq where
\bea
Q_{cl}(N,T)=\frac{\Gamma(1+1/N)}{\sqrt{\pi}\beta^{1/2+1/N}}
\eea
is the semiclassical partition function
corresponding to the Hamiltonian $p^{2}+|x|^{N}$ \cite{mar84}.
As $Z_{cl}$ and $Q_{cl}$ are proportional to each other,
they lead to the same thermodynamics as $T\rightarrow \infty$
(see plots), and in particular the specific heat is the same
in both cases. Nevertheless there is an important
difference, namely, the characteristic temperature associated with
$Q_{cl}$ is bounded [it is given by $\Theta$ in (\ref{ctemperature})
omitting $\sin(M\pi/N)$],
and consequently their corresponding thermodynamic functions
may differ significantly at intermediate and small temperatures.
For example, if $T\ll\Theta$, the corresponding specific heat would have
the usual quantum behavior expressed by an exponential decay
as $T\rightarrow 0$,
\begin{equation}
C\simeq \left(\frac{\Theta}{T}\right)^{2}e^{-\Theta/T},
\label{qsheat}
\end{equation}
whereas the same $T$ might be much greater than  the
characteristic temperature associated with $Q_{cl}$,
for which the classical behavior in (\ref{sheat}) would be observed.
[Strictly speaking, $\Theta$ in (\ref{qsheat})
should be replaced by the energy gap between the first excited
state and the ground state, but this inaccuracy does not spoil the
argument.]
\section{Classical statistical mechanics}
\label{cmechanics}
\renewcommand{\thefootnote}{\dag}
\setcounter{footnote}{1}
In standard classical statistical mechanics, with a Hermitian Hamiltonian,
the partition function is given by integrating over the real phase space,
\bea\label{zcla}
 Z_{cl}(T)&=&\frac{1}{2\pi}\int_{-\infty}^{+\infty}dp
\int_{-\infty}^{+\infty}dx\ e^{-\beta H(p,x)},\nonumber\\
&=&
\frac{1}{2\sqrt{\pi\beta}}
\int_{-\infty}^{+\infty}dx\ e^{-\beta V(x)}
\eea
for $H=p^2+V(x)$.

For the non-Hermitian Hamiltonians of (\ref{M<>1}) it is not immediately clear what is the correct
formulation. For sufficiently small $\epsilon$ the integral of (\ref{zcla}) is still convergent, and indeed
reproduces correctly the high-temperature limit of (\ref{zq}). However, for larger values of $\epsilon$, for
example $\epsilon>1$ in the case $M=1$,  the integral along the real $x$-axis diverges, and in order to obtain a
convergent result the contour of integration must be continued into the lower half of the complex $x$-plane.

The situation is similar, but not identical, to the Stokes wedges encountered in the quantum problem~\cite{BB}.
Thus, let us set $x=r e^{-i\theta}$. The condition for convergence for large $r$ is that ${\rm Re}\ V>0$,
with $V=r^Ne^{i(\pi\epsilon/2-N\theta)}$,
leading to
\beax
\cos\left[N\left(\frac{\pi}{2}-\theta\right)-M\pi\right]>0\ .
\eeax
This leads to wedges in the complex $x$-plane where the integral is convergent and other, forbidden, wedges where it is not.
The forbidden wedges rotate downwards and become narrower as $N$ increases, in a similar manner to Stokes wedges.

\begin{figure}[thp]
\begin{center}
\includegraphics[scale=0.8]{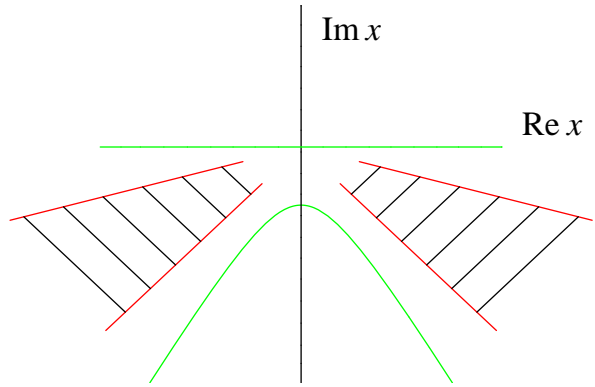}
\caption{The classically forbidden wedges (hatched) for $N=6$. The hyperbolic contour
is a possible contour for the non-Hermitian $V=x^6$ ($M=1$, $\epsilon=4$), whereas integration along
the real axis corresponds to $V=|x|^6$ ($M=3$, $\epsilon=0$). }
\label{wedge}
\end{center}
\end{figure}

In the case $M=1$, i.e. continuing away from the harmonic oscillator, the right-hand forbidden wedge is centered on $\theta_N=(\pi/2)(1-4/N)$
and has opening angle $\pi/N$. There is a mirror image in the left-half plane. So for $N=3$ the lower edge of the forbidden wedges
lies precisely on the real axis, confirming that $ix^3$ is a limiting case. For $N=4$, the forbidden wedges include the real axis, which is no longer viable as an
integration contour. When $N=5$ the upper edge of the wedges lies on the real axis, and thereafter the wedges lie entirely
in the lower half plane. This latter situation is illustrated in Fig.~\ref{wedge} for $N=6$. In this case the real axis is again
a viable contour, but in fact it corresponds to the Hermitian potential $V=|x|^6$, whereas the hyperbolic contour is a possible
contour for the non-Hermitian theory: it has been pushed off the real axis by the continuation in $\epsilon$, and the two contours
are separated by a forbidden region. There is in fact another forbidden wedge centered on the negative imaginary axis, which
is avoided by the contour shown.

Moreover, these integrals are not only convergent, but they correctly reproduce the semiclassical result of (\ref{zq}).
Thus, for example, in the case $N=6$, if we integrate along the real axis we obtain $Q_{cl}(6, T)$, corresponding to
the Hermitian case $M=3$, $\epsilon=0$, but if we integrate along the rays $x=\pm\ r \exp(\mp\ i\pi/3)$, the centres of the next allowed left and right wedges, we obtain precisely the extra factor $\cos(\pi/3)=\sin(\pi/N)$ required to give $Z_{cl}$ for $M=1$.

In principle, an alternative way of determining what is the correct expression for the classical partition function is
to start with the definition in terms of the equivalent Hermitian Hamiltonian and make the appropriate changes of variables.

As mentioned in the Introduction, in quantum mechanics a non-Hermitian Hamiltonian $H$ with a completely real spectrum is related to an isospectral Hermitian
Hamiltonian $h$ by the similarity transformation~\cite{AM02}:
\bea\label{sim}
H=\rho^{-1} h \rho\equiv e^{\half Q} h\, e^{-\half Q},
\eea
where the metric $\eta_+$ is written~\cite{BBJ} in terms of the Hermitian operator $Q$ as $\eta_+=e^{-Q}$.
The operator $\rho$ is therefore Hermitian and positive definite.
Thus we can certainly write $Z_{cl}$ as
\begin{equation}
Z_{cl}(T)=\frac{1}{2\pi}\int\!\!\!\int_{-\infty}^\infty dx\, dp\ e^{-\beta h(x,p)}.
\label{cpartition}
\end{equation}
From (\ref{sim}) we have the condition of quasi-Hermiticity on $H$:
\bea\label{qH}
H^\dag=e^{-Q} H e^Q\equiv \eta H\, \eta^{-1}.
\eea
Similarly any observable $A$, with real expectation values, must also be quasi-Hermitian, i.e.
\bea\label{qA}
A^\dag=e^{-Q} A e^Q\equiv \eta A\, \eta^{-1}.
\eea
This in turn means that $A$ is related to a standard Hermitian counterpart $a$ by the same similarity transformation:
\beax
A=e^{\half Q} a\, e^{-\half Q}.
\eeax
We can use this latter equation to derive two different relations between $H$ and $h$, namely
\bea\label{identities}
H(x,p)&=&e^{\half Q}h(x,p)e^{-\half Q}=h(X,P),\nonumber\\ && \\
H(X^\dag,P^\dag)&=&e^{\half Q}h(X^\dag,P^\dag)e^{-\half Q}=h(x,p),\nonumber
\eea
It is the second of these identities that we need here, to write
\beax
Z_{cl}(T)=\frac{1}{2\pi}\int\!\!\!\int_{-\infty}^\infty dx\,dp\ e^{-\beta H(X^\dag,\ P^\dag)}.
\eeax

Now make the change of variables to $\xi\equiv X^\dag$,\ $\pi\equiv P^\dag$, to be treated here as classical variables.
Thus, in the first instance (the Jacobian is 1),
\bea\label{Zcontour}
Z_{cl}(T)=\frac{1}{2\pi}\int\!\!\!\int_C d\xi\,d\pi\ e^{-\beta H(\xi,\ \pi)},
\eea
where $C$ is a contour in complex $(\xi, \pi)$ space determined parametrically by $\xi=X^\dag(x,p)$,
$\pi=P^\dag(x,p)$. That is, the initial contour is ultimately determined by the metric $\eta$.  Whether or not the contour can
subsequently be deformed to a standard real phase space $(\xi, \pi)$ is a matter to be determined for an individual Hamiltonian.
From our preceding discussion it seems clear that this is not possible in general because of the presence of intervening
forbidden wedges.

Note that the metric $\eta$ does not appear explicitly as an integration measure in the integral representation (\ref{Zcontour}) of the partition function, only through the form of the relationship between ($\xi$,\ $\pi$) and ($x$,\ $p$). The issue of the role of $\eta$ in path integrals has been addressed and clarified in \cite{JR, AM}.

As mentioned above, there are very few cases where $Q$, and hence the relations between ($x$,\ $p$) and ($\xi$,\ $\pi$), are known exactly. Apart from the rather trivial case of the Swanson model~\cite{MS} (where in fact standard phase space can be used), a nice example~\cite{JM} within the class of Hamiltonians of (\ref{M<>1}) is the wrong-sign quartic $V=-x^4$.

A word of clarification is in order here. The original Hamiltonian
in this case is really $H_z(z,p_z)=p_z^2-z^4$, where the
eigenvalue problem has to be posed on a complex contour in the
appropriate Stokes wedges. It is only when a particular contour is
chosen, with the parametrization $z=-2i\surd(1+ix)$ in terms of
the real variable $x$, that we obtain the non-Hermitian
Hamiltonian $H(x,p)$ for which $Q$ was found, namely
\beax
H(x,p)=\half\left\{1+ix, p^2\right\}-\half p - \alpha (1+ix)^2,
\eeax
where, in the present case, $\alpha=16$. The $Q$ operator is
\beax
Q=-\frac{p^3}{3\alpha}+2p,
\eeax which results, via the
first equation of (\ref{identities}), in the equivalent Hermitian Hamiltonian
\beax
h(x,p)=\frac{p^4}{4\alpha}-\half p +\alpha x^2.
\eeax
This is still not a conventional Hamiltonian, but becomes so on taking the
Fourier transform.

It is easily verified that using this $h(x,p)$ in (\ref{cpartition}) correctly gives the appropriate result
($M=1$, $\epsilon= 2$) in (\ref{zq}). This is true whether or not one includes the linear term $-p/2$,
which in fact is a quantum anomaly~\cite{ben06} proportional to $\hbar$.

If we now make the transformation to the variables $X^\dag=\xi$ and $P^\dag=\pi$, using the second equation of (\ref{identities}),
we obtain
\bea\label{para}
&&\pi=p\nonumber\\
&&\xi=x+i\left(1-\frac{p^2}{2\alpha}\right)\\
&&H(\xi,\pi)=\half\{1+i\xi,\pi^2\}-\half\pi-\alpha(1+i\xi)^2\nonumber.
\eea
We now have an expression for $Z_{cl}$ of the form of (\ref{Zcontour}), in which the contour is given
parametrically by (\ref{para}) in terms of the real variables $x$ and $p$. However, it is readily verified
that there is no obstruction to deforming the contour to the real axis, so that $Z_{cl}$ can be expressed
as a real phase-space integral. But $H(\xi,\pi)$ has precisely the same form as $H(x,p)$, and we can take the reverse step
to the original variables $z$ and $p_z$. That is,
\bea
z&=&-2i\surd(1+ix)\nonumber\\
p_z&=& p\surd(1+ix),
\eea
a canonical transformation from $(x,p)$ to $(z, p_z)$. For any finite $x$, the argument of the variable $p_z$
is less in modulus than $\pi/4$, so that it lies within the wedge including the real axis that guarantees convergence of the integral $\int dp_z \exp(-\beta p_z^2)$. Hence the $p_z$ integral can be deformed to the real axis. However, for the $z$ integration ($N=4$) the contour cannot be so deformed because of an intervening forbidden wedge. This is in agreement
with the results found previously when we performed the integrals over appropriately chosen rays, to obtain the
correct semiclassical result of (\ref{zq}).
\section{Discussion}
\label{disc}

To summarize, in this work we addressed the quantum and
classical statistical mechanics of the class of non-Hermitian $PT$-symmetric oscillators of (\ref{M<>1}). These are in principle related by a similarity transformation to an equivalent Hermitian Hamiltonian $h$. However, for the quantum partition function $Z(T)$ one only needs the energy levels of $H$, which we evaluated using the WKB approximation including the first subleading correction. The main qualitative difference from the Hermitian oscillators ($M=N/2$) with potential $|x|^N$ turns out to be that because of the factor $\sin(M\pi/N)$ in the denominator the characteristic temperature $\Theta(M,N)$ of (\ref{ctemperature}) grows without limit as $N$ increases, so that the onset of semiclassical behavior is progressively delayed. At high temperatures the entropy has the same form as that for the $|x|^N$ oscillator, except for the different value of $\Theta$, and the specific heat is the same, depending only on $N$.

The semiclassical partition function $Z_{cl}(T)$ was determined in the first instance as the high-temperature limit of $Z(T)$. We then investigated how this result could be reproduced by a purely classical calculation, and found that
this could be achieved only by extending the classical phase-space integrations into the complex plane. Specifically, because
of the simple dependence of $H$ on $p$, the $p$ integration can remain on the real axis, but the $x$-dependence means that the
$x$ integration can only be done within certain allowed wedges in the complex $x$ plane, corresponding to different values of $M$. By integrating along the rays at the centre of the allowed wedges we verified that we indeed reproduced the semiclassical result
of (\ref{zq}). In order to understand this from an alternative perspective we considered the special case of the ``upside-down" quartic, the one example of the class (\ref{M<>1}) where $h$ is known explicitly. Starting with the conventional Hermitian formulation for $Z_{cl}$ in terms of $h$ we implemented the similarity transformation to re-express it in terms of $H$, showing that indeed it required a complex contour in $x$ of the type we had previously found.

It has been shown \cite{Smilga} that the classical Hamiltonian dynamics for systems of the type (\ref{M=1}) and (\ref{M<>1})
can be formulated in a real 4-dimensional phase space\red{, while a generalized canonical structure for non-Hermitian classical
dynamics has recently been derived in \cite{E-M}.} It would be interesting to see how our results for the semiclassical
partition function can be derived in those formalisms.

\ack{
ESM is grateful to Prof.~Chris Hull (Theory Group at Imperial College London) for the hospitality.
ESM was partially supported by the research agency CAPES.}

\end{document}